\documentclass[referee,pdflatex,sn-mathphys-num]{sn-jnl}

\usepackage{graphicx}%
\usepackage{multirow}%
\usepackage{amsmath,amssymb,amsfonts}%
\usepackage{amsthm}%
\usepackage{mathrsfs}%
\usepackage[title]{appendix}%
\usepackage{xcolor}%
\usepackage{textcomp}%
\usepackage{manyfoot}%
\usepackage{booktabs}%
\usepackage{algorithm}%
\usepackage{algorithmicx}%
\usepackage{algpseudocode}%
\usepackage{listings}%
\usepackage{subcaption}
\usepackage{units}%
\usepackage{float}%

\theoremstyle{thmstyleone}%
\theoremstyle{thmstyletwo}%

\theoremstyle{thmstylethree}%

\begin{document}

\title[Exclusive Dyon]{Exclusive Dyon Production in High-Energy Collisions}

\author{\fnm{Luis} \sur{F. C. In\'{a}cio}}\email{lfci.fisica@gmail.com}

\author*{\fnm{Werner} \sur{K. Sauter}}\email{werner.sauter@ufpel.edu.br}

\affil{\orgdiv{Instituto de F\'{i}sica e Matem\'{a}tica}, \orgname{Universidade Federal de Pelotas}, \orgaddress{\street{Campus Universit\'{a}rio do Cap\~{a}o do Le\~{a}o}, \city{Pelotas}, \postcode{96001-970}, \state{Rio Grande do Sul}, \country{Brasil}}}


\abstract{In this work, we investigate the exclusive central production of dyons, a particle carrying electric and magnetic charges, in hadronic interactions at LHC energies, assuming the photon fusion mechanism. Motivated by predictions from theoretical physics beyond the Standard Model, we analyze the photoproduction of these particles in the ultraperipheral collision regime, employing the equivalent photon approximation. We estimate the cross-sections for dyons production in spin 0, 1/2, and 1 scenarios, adopting values for electric and magnetic charges from the literature. Our results demonstrate a direct correlation between spin and cross section, predicting a substantially higher probability for the photoproduction of higher-spin dyons.}

\keywords{Exotic Particles, Dyons, Ultraperipheral Collisions, Photon Fusion.}



\maketitle

\section{Introduction}
\label{intro}

The exploration of physics beyond the Standard Model (BSM) has increasingly focused on the search for exotic hypothetical particles, propelled by the extreme collision energies available at the LHC. While direct experimental evidence for these states remains elusive, they are strongly motivated by advanced theoretical frameworks seeking to resolve fundamental shortcomings and unanswered questions in modern particle physics.

Among the several predicted BSM particles, there is the dyon, an exotic particle that is highly ionizing and have both electrical and magnetic charges. The first hypothesis of the dyon existence was proposed by Julian Schwinger in 1969 within the context of a unified theory for matter, which he used to construct what he called the magnetic model of matter \cite{Schwinger:1969ib}. Using this model, Schwinger predicted the existence of a particle with characteristics similar to those of the $J/\psi$ particle, which was experimentally detected in 1975.

The magnetic monopole as source of the magnetic field is still widely discussed in Physics (for a recent review, see \cite{Mavromatos:2020gwk}). With the advent of quantum mechanics, Dirac in 1931 studied magnetic mono\-poles \cite{Dirac1931}  and demonstrated that the existence of this source of the magnetic field led to the quantization of electric charge. Since then, extensive research has been conducted in search of this particle. Schwinger's original work has shown that dyons arise generically in theories with mono\-poles, including many particle physics theories, such as the Grand Unified Theory (GUT) \cite{Dyon_GUT82,Dyon_GUT84}, Einstein-Yang-Mills Theory \cite{Dyon_Einstein-Yang-Mills2000,Dyon_Einstein-Yang-Mills2012}, Kaluza-Klein Theory \cite{Dyons_Kaluza-Klein1997}, String Theory \cite{Dyons_Kaluza-Klein2011}, and M Theory \cite{Dyons_teoriaM2001}. The experimental detection of dyons would help answer very important questions related to any of the theories mentioned above, and would also provide crucial information about the source of the magnetic field.

\begin{figure}[htb]
	\centering
	\resizebox{0.4\textwidth}{!}{\includegraphics{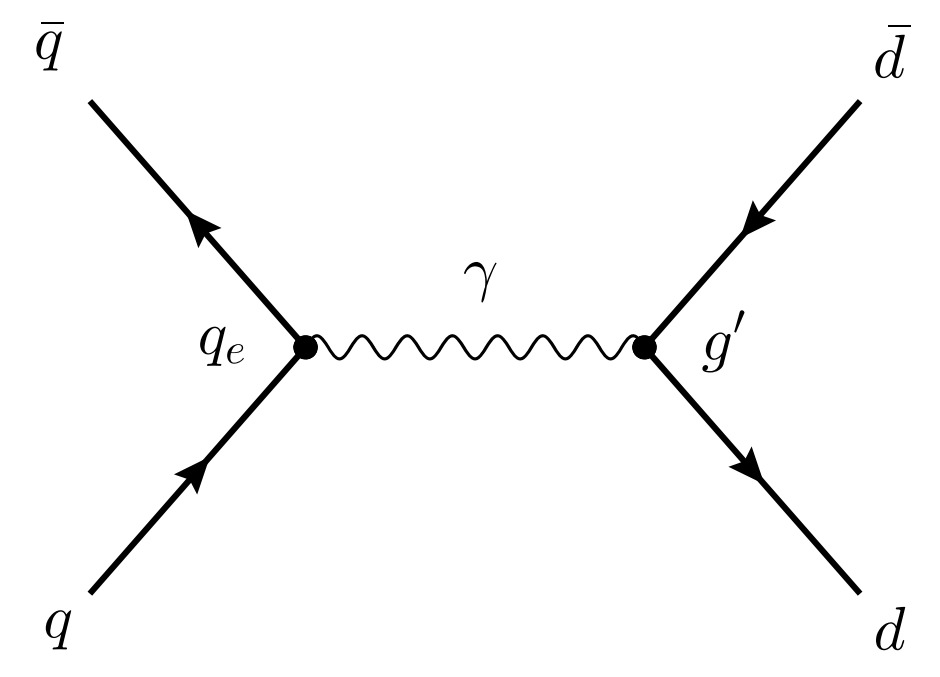}}
	\caption{Feynman diagram for the production of the dyon-antidyon pair through the Drell-Yan production mechanism.}
	\label{fig:Drell_Yan}       
\end{figure}

Currently, the experiment designed to search for these exotic hypothetical particles is the Monopole and Exotics Detector at the LHC (MoEDAL\-)\cite{MoEDAL:2014ttp,MoEDAL1}. During the second phase of the LHC operation (Run-2), between 2015 and 2017, the MoEDAL detector performed the first search for dyons in proton-proton collisions with an energy of \unit[13]{TeV} \cite{primeirabusca}. In this search, the collaboration used the Drell-Yan (DY) process as the production mechanism, in which a quark and an antiquark annihilate, producing a photon that, in turn, generates a pair of leptons, in this case, a dyon and an antidyon (see Figure~\ref{fig:Drell_Yan}). The collaboration estimated the production of a particle carrying a magnetic charge that varied up to six units of a fundamental magnetic charge (the Dirac charge) and an electric charge of up to 200 times the electron charge for dyons with masses between 750 and \unit[1910]{GeV}. However, no dyons were detected by the collaboration within the experimental cutoffs and limits considered in the article above, motivating the exploration of a different production mechanism, the photon fusion.

In this work, we analyze the exclusive central production of dyons through the photon fusion mechanism in ultraperipheral collisions of ultra-relativistic charged particles (proton-proton) at energies available at the LHC. Exclusive central production is an useful process for searching and studying exotic particle states, as the projectiles do not dissociate and the particle is produced in the central rapidity region of the detector, providing a clean experimental signal of this process. While the MoEDAL collaboration established exclusion limits for dyons produced via the DY mechanism, those results are model-dependent and do not encompass all possible production channels. Our statement that current results are inconclusive refers to the lack of direct experimental observation in alternative channels, such as photon fusion, which offers a distinct experimental signature and can probe regions of the parameter space where DY production might be suppressed.

\section{Formalism}
\label{sec:1}

The formalism adopted in this work for the exclusive production of dyons is the Equivalent Photon Approximation in the ultraperipheral collisions (for a review, see \cite{bertulani2005physics, baltz2008physics}). Through this formalism, the total cross section for the collision of two charged particles can be expressed as the product of the pair photoproduction cross section and the equivalent photon spectrum, characteristic of each charged particle. The photon spectrum was then integrated over the entire range of photon energies, providing an accurate estimate of the dyon production rate in ultraperipheral proton-proton collisions at the LHC. The expression for the cross section as a function of center-of-mass energy $s$ is given by
\begin{equation}\label{2.1}
    \sigma_{pp \to p \mathcal{X} p}(s) = \int^1_{4m^2/s} dx_1 \int^1_{4m^2/x_1s}  dx_2 \, f(x_1) f(x_2) \, \sigma_{\gamma\gamma \to \mathcal{X}}(W_{\gamma\gamma})\,,
\end{equation}
where \( x_1 \) and \( x_2 \) are the ratio of energies of the photons and projectiles, \( f(x_1) \) and \( f(x_2) \) are the equivalent photon spectra emitted by the hadrons, \( W_{\gamma\gamma} = x_1x_2s \) is the center-of-mass energy of the two photons, and \( \sigma_{\gamma\gamma \longrightarrow \mathcal{X}}(W_{\gamma\gamma}) \) is the cross section for the production of state \( \mathcal{X} \) through the fusion of two photons, It will be calculated analytically in tree order for different spins, as presented in Section (\ref{sec:2}).

We considered the production of dyons in hadron collisions. It is well known that the proton is a particle with a internal structure, composed of quarks and gluons that strongly interact with each other. To model the proton charge distribution, we use the expression for the photon energy spectrum derived by Drees and Zeppenfeld \cite{Drees_Zeppenfeld1989}. The choice of this particular parametrization is justified by its accurate treatment of the protons non-point-like nature. This approach is critical for ultraperipheral collisions (UPC), as it accounts for the electromagnetic form factors and the $Q^{2}$ dependence of the emitted virtual photons, thereby reducing theoretical uncertainties in the equivalent photon spectrum. The expression of photon flux is given by
\begin{equation}\label{2}
	f(x) = \frac{\alpha_{el}}{\pi}\frac{1-x+0.5x^2}{x}\left[ \ln(A) - \frac{11}{6} + \frac{3}{A} - \frac{3}{2A^2} + \frac{1}{3A^3}\right].
\end{equation}
where
\begin{equation}
	A = 1 + \dfrac{\unit[0.71]{GeV^2}}{Q_{min}^2}, \qquad Q_{min}^2 = \dfrac{m^2_p x^2}{1-x},
\end{equation}
with \( Q^2 \) the four-momentum transferred from the projectile, \( \alpha_{el} \) the electromagnetic coupling constant, \( m_p \) the mass of the proton.

In the following, some points about the formalism are discussed. While UPCs are characterized by impact parameters larger than the sum of the nuclear radii, strictly isolating pure QED processes in proton-proton collisions requires careful consideration of strong interaction backgrounds. In standard central exclusive production, diffractive mechanisms, such as double, Pomeron exchange (DPE), can mimic the kinematic signature of photon fusion. However, we justify the dominance of the photon fusion channel for dyon production on two grounds. First, assuming the BSM dyons investigated in this effective framework are color-singlet states, their direct tree-level production via gluon-gluon fusion is inherently suppressed. Second, the extremely large effective coupling $\alpha_d$, driven by the fundamental magnetic charge ($g_D$), massively enhances the $\gamma\gamma \to d\bar{d}$ cross section, making the electromagnetic channel overwhelmingly dominant over potential QCD background processes.

Furthermore, it is crucial to emphasize that the EPA formulation yields the bare theoretical cross sections. In realistic LHC operational scenarios, soft multiparton rescattering between the interacting protons can produce secondary hadrons that populate the detector's central rapidity region, effectively destroying the exclusive signature. To obtain the observable experimental cross section, our theoretical estimates would need to be multiplied by a rapidity gap survival probability factor ($S^2$). For two-photon processes in $pp$ collisions, $S^2$ is generally high compared to QCD-mediated diffraction, but strictly less than unity. Consequently, the cross sections and expected event yields reported in this work omit these absorptive corrections and should be interpreted as theoretical upper bounds, establishing a phenomenological baseline for future detector-level simulations.


\section{Photoproduction of Dyons}
\label{sec:2}

As mentioned earlier, in the first search attempt for dyons, the chosen production mechanism was Drell-Yan, based on investigations of magnetic monopoles with spins 0, 1/2, and 1 conducted in the MoEDAL experiment, as documented in \cite{mono2} and \cite{mono1}. The models were generated using MadGraph5 \cite{Alwall:2014hca} with the universal FeynRules output, as described in \cite{mono2}. Additionally, in this initial search for dyons in MoEDAL, tree-level diagrams and NNPDF23 Parton Distribution Functions, as described in \cite{Ball2012}, were used for the DY process.

\begin{figure}
	\centering
	\resizebox{0.4\textwidth}{!}{\includegraphics{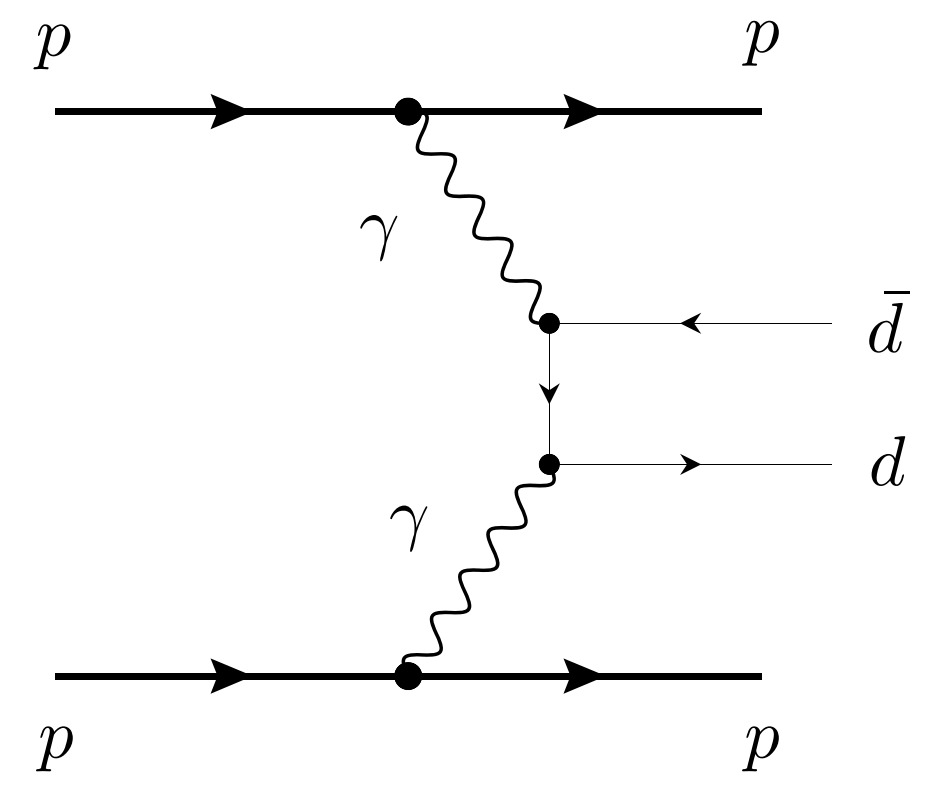}}
	\caption{Feynman diagram for the production of the dyon-antidyon pair through the Photon Fusion production mechanism.}
	\label{fig:fotoproducao}       
\end{figure}

In this study, we explore another mechanism: the photon fusion, where two high-energy photons emitted by charged particles interact and produce particles \cite{ennadifi2024light}. Our interest lies in the production of a dyon-antidyon pair, as illustrated in Figure (\ref{fig:fotoproducao}). In other words, the photon fusion production mechanism is associated with a quantum transition, in which two photons transform into two other particles. This transition can result in the creation of a fermion (such as leptons or quarks) or boson pair (such as $W^{+}W^{-}$ or $Z^{0}$) or a bound state \cite{dEnterria2013}. The charged nature of the dyons ensures that they can be produced by photon fusion. For a discussion of photon physics at the LHC and realistic prospects for future colliders, see \cite{klasen2019perspectives}.

Photon fusion is only possible when the photons have sufficient energy to produce the desired particle masses. In practice, photon fusion is uncommon at low energies owing to its low probability. However, photon fusion can occur in high-energy environments, such as particle collisions in accelerators. The high-energy environment of LHC favors these interactions, making photon fusion an effective tool for exploring new frontiers in particle physics and the production of hypothetical particles, as seen in  \cite{belforte2019evidence,harland2016production,da2015central,d2022observing,chiesa2024multi,vidovic1993impact,csahin2009probe,dougall2009dirac,szczurek2002heavy, Bertulani:2024vpt}. The collision energy limit adopted in this work is \unit[14]{TeV}, which corresponds to the limit to be reached by the LHC in its third operational phase (Run-3). For a review of this production mechanism, see  \cite{Mikaelian1981, Moulin1996, Jikia1993, dEnterria2008p}.

In particular, a dyon is a particle with a unique signature, unlike magnetic monopoles, primarily because of its electric charge. The theoretical approximations developed for monopoles in \cite{mono2} can naturally be extended to include dyons. For this, it is necessary to replace, in the expression for the cross section, the square of the monopole's magnetic charge, \( g^2 \), by \( g^2 + q^2 \), where \( q \) represents the electric charge of the dyon. This modification is consistent with the dual effective theory of Milton and Gamberg, as described in \cite{Gamberg2000} and \cite{Gamberg:2000}. This approach incorporates the additional electromagnetic contribution of dyons while maintaining coherence with the established theoretical framework for monopoles.

According to \cite{primeirabusca}, the formalism can be extended to include the photoproduction of dyons. The electromagnetic fine-structure constant is given by
\begin{equation} \label{1}
	\alpha_{el} = \frac{e^2}{4\pi} \approx \frac{1}{137},
\end{equation}
through the process of electromagnetic dualization, we derive from (\ref{1}) the magnetic fine-structure constant
\begin{equation} \label{4.2}
	\alpha_{g} = \frac{g^2 (\beta)}{4\pi} = \frac{g^2 \beta^{2\delta}}{4\pi} = \alpha_{g}^2 \beta^{2\delta},
\end{equation}
where $\delta$ is a phenomenological parameter governing the velocity dependence of the magnetic coupling. Following standard monopole effective models, we adopt $\delta = 1$, which recovers the familiar $\beta^2$ dependence in the coupling, as applied in subsequent equations.

To obtain the fine-structure constant for the dyons, \( \alpha_{d} \), it is sufficient to substitute in (\ref{4.2}), $g^2$ by $g^2 + q^2$, as shown below,
\begin{equation} \label{4.3}
	\alpha_{d} = \frac{\left( g^2 + q^2\right) \beta^2}{4\pi},
\end{equation}
where \( g = n_m g_D \), \( n_m \) is the magnetic charge number and \( g_D \) is the fundamental magnetic charge (\( g_D = 68.5e \)). And also \( q = n_e e \), where \( n_e \) is the electric charge number and \( e \) is the fundamental electric charge. Thus, we have:
\begin{equation} \label{4.4}
	\alpha_{d} = \frac{\left( n_m^2 g_D^2 + n_e^2 e^2\right) \beta^2}{4\pi}.
\end{equation}

For dyons, the fine-structure constant is generalized to include both electric and magnetic interactions. After some mathematical manipulations in (\ref{4.4}), the fine-structure constant can be written as
\begin{equation} \label{4.5}
	\alpha_{d} = \beta^2\alpha_{el}\left[\left(\dfrac{n_m}{2\alpha_{el}}\right)^2  + n_e^2\right].
\end{equation}

We are using a model dependent on $\beta$, measured in the photon-photon center-of-mass frame. This parameter is kinematically defined as
\begin{equation} \label{4.6}
	\beta = \sqrt{1 - \frac{4M_d^2}{s_{\gamma\gamma}}},
\end{equation}
where $M_d$ is the dyon mass and $s_{\gamma\gamma}$ represents the invariant mass squared of the interacting two-photon system. 

We used the model of electromagnetic interactions of a magnetic monopole with spins $0$, $1/2$, and $1$ with ordinary photons developed in \cite{Baines:2018ltl}. The corresponding theory is an effective $U(1)$ gauge theory obtained after the appropriate dualization of the relevant field theories that describe the interactions of charged spin fields with photons. In reference \cite{kurochkin2006}, the authors calculated (in suitably dualized models) the total cross sections for monopole pair production via photon fusion for three different spin models: spins $0$, $1/2$, and $1$. Theoretically, the cross sections increase with increasing spin, given a fixed particle mass. 
Based on this work, we can generalize the cross section to the case of dyons with spin 0, 1/2 and 1 adding the eletric charge and modifing the coupling as done above.

We will use the cross sections from \cite{kurochkin2006}, generalizing to the case of dyons. As in the case of the production of magnetic monopoles \cite{Reis:2017rvb, daSilva:2023jxd}, due the large coupling, a full perturbative calculation cannot be done and the scattering amplitudes are calculated in lowest (tree) order. Therefore, the differential cross section for the photoproduction of the spin $0$ dyon-antidyon pair is given by
\begin{equation} \label{4.7}
	\frac{\text{d} \sigma_{\gamma \gamma \to d\bar{d}}^{S=0}}{\text{d} \Omega} = \frac{ \alpha_d^2 (\beta)\beta}{2 s_{\gamma\gamma}} \left[1 + \left(  1 - \frac{2\left( 1 - \beta^2\right) }{1 - \beta^2 \cos^2 \theta}  \right)^2 \right],
\end{equation}
and the total cross section becomes
\begin{equation} \label{7}
	\sigma_{\gamma \gamma \to d\bar{d}}^{S=0} = \frac{4\pi \alpha_d^2 (\beta)\beta}{s_{\gamma\gamma}} \left[2-\beta^2 - \dfrac{1}{2\beta} (1-\beta^4) \ln{\left(  \dfrac{1+\beta}{1-\beta}   \right)}\right],
\end{equation}
where $s_{\gamma\gamma}$ is the center-of-mass energy of photon fusion.

For the production of a dyon-antidyon pair with spin $1/2$, the differential cross section for the photoproduction of the pair is given by
\begin{multline}
	\frac{\text{d}\sigma_{\gamma \gamma \to d\bar{d}}^{S=\frac{1}{2}}}{\text{d}\Omega} = \frac{\alpha_d^2 (\beta)\beta}{4s_{\gamma\gamma}(1 - \beta^2\cos^2\theta)^2}
	\\
	\times\Big(-8\beta^4 + 8\beta^2 - 4\beta^4\cos^4\theta + 8\beta^4\cos^2\theta - 8\beta^2\cos^2\theta + 4    \Big),
\end{multline}
and the total cross section becomes
\begin{equation}\label{4.11}
	\sigma_{\gamma\gamma \to d\bar{d}}^{S=\frac{1}{2}} = \frac{2\pi\alpha_d^2(\beta)}{s_{\gamma\gamma}} \left[ (3 - \beta^4)\ln\left(\frac{1+\beta}{1-\beta}\right) - 2\beta(2 - \beta^2) \right].
\end{equation}

Finally, the differential cross section for the photoproduction of the dyon-antidyon pair with spin $1$ via photon fusion, is given by
\begin{multline}
	\frac{\text{d}\sigma_{\gamma \gamma \to d\bar{d}}^{S=1}}{\text{d}\Omega} = \frac{\alpha_d^2 (\beta)\beta}{2s_{\gamma\gamma}(1 - \beta^2\cos^2\theta)^2}
	\\
	\times\Big(3\beta^4(\cos^4\theta - 2\cos^2\theta + 2)  + \beta^2(16\cos^2\theta - 6) + 19   \Big),
\end{multline}
and the total cross section becomes
\begin{equation}\label{4.13}
	\sigma_{\gamma \gamma \to d\bar{d}}^{S=1} = \frac{\pi \alpha_d^2 (\beta)\beta}{s_{\gamma\gamma}} \left( 2 \frac{3\beta^4 - 9\beta^2 + 22}{1-\beta^2} - \frac{3(1-\beta^4)}{\beta} \ln{\left(\frac{1+\beta}{1-\beta}\right)} \right).
\end{equation}

\section{Results}

In this section, we present the calculations for the production of dyon-antidyon pairs considering $\gamma\gamma$-interactions in proton collisions with a center-of-mass energy of \( \sqrt{s} = \unit[14]{TeV} \) at the LHC. For this, we analyze dyons with different values of electric and magnetic charges, and compare the resulting cross sections for dyons with spin $0$, $1/2$ and $1$.

\begin{figure}[H]
	\centering
	\resizebox{0.5\textwidth}{!}{\includegraphics{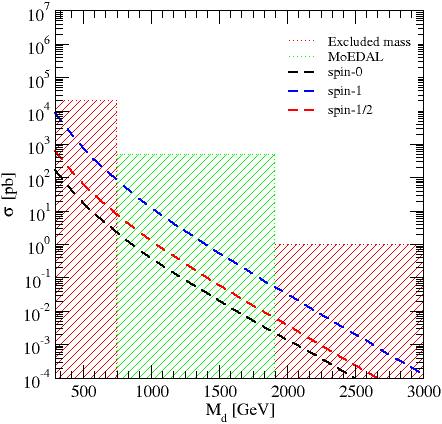}}
	\caption{Cross section for the production of the dyon-antidyon pair via photon fusion for the three types of spin, with magnetic charge \(1g_D\) and electric charge \(1e\), in proton-proton collisions with a center-of-mass energy of \unit[14]{TeV}.}
	\label{fig:11}       
\end{figure}

For dyons with $n_e=1, n_m=1$, the behavior of the cross sections (\ref{7}), (\ref{4.11}), and (\ref{4.13}) as a function of mass ($M_d$) for the production of dyon-antidyon pairs via photon fusion is shown in Figure~(\ref{fig:11}). The evaluated parameter space is overlaid by the existing experimental constraints, dividing between exclusion regions (highlighted in red) and the viable mass window (highlighted in green). The graph displays the cross sections calculated for the three types of spin ($0$, $1/2$ and $1$).

The region on the left ($M_d < \unit[750]{GeV}$) represents the parameter space previously discarded by experimental collaborations. As illustrated by the asymptotic behavior of the curves, the theoretical cross-section for lower masses reaches significantly high values. The absence of any excess events recorded by the detectors in the face of this high expected production rate excludes the existence of dyons in this mass range.

In contrast, the central band ($750 \leq M_d \leq \unit[1910]{GeV}$) delineates the ``allowed window'' of the theory \cite{primeirabusca}. This zone defines the parameter space that is still unexplored or where current experimental sensitivity is not sufficient to refute the model. In this region, a clear hierarchy in production rates governed by the particle spin is observed, where $\sigma_{\gamma \gamma \to d\bar{d}}^{S=1} > \sigma_{\gamma \gamma \to d\bar{d}}^{S=1/2} > \sigma_{\gamma \gamma \to d\bar{d}}^{S=0}$. This behavior indicates that vector dyons (spin-1) exhibit the most prominent phenomenological signature.

Finally, the region on the right ($M_d > \unit[1910]{GeV}$) indicates the upper limit of current searches. Due to severe suppression of the phase space at the high-mass limit, the probability of production decays exponentially for values that are below the statistical observability threshold of the collider. Therefore, the absence of discoveries above $\unit[1910]{GeV}$ reflects purely kinematic and luminosity limitations of the experiment, outlining the sensitivity frontier of the current generation of detectors.

\begin{figure}[H]
     \centering
     \begin{subfigure}[b]{0.45\textwidth}
         \centering
         \includegraphics[width=\textwidth]{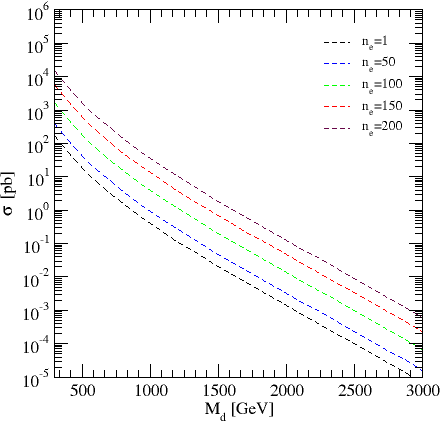}
         \caption{Spin 0}
         \label{fig:spin-0}
     \end{subfigure}
     \hfill
     \begin{subfigure}[b]{0.45\textwidth}
         \centering
         \includegraphics[width=\textwidth]{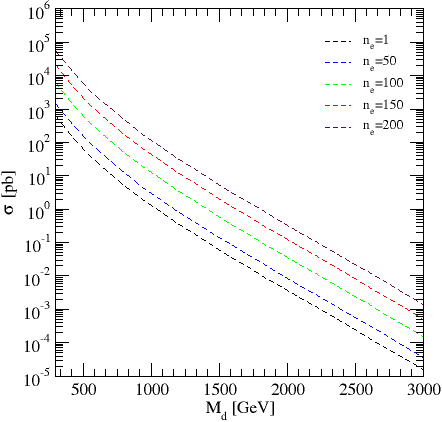}
         \caption{Spin $1/2$}
         \label{fig:spin-1/2}
     \end{subfigure}
     \hfill
     \begin{subfigure}[b]{0.45\textwidth}
         \centering
         \includegraphics[width=\textwidth]{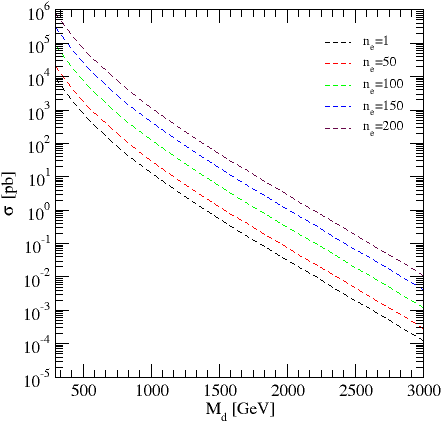}
         \caption{Spin 1}
         \label{fig:spin-1}
     \end{subfigure}
     
     \caption{Behavior of the cross section for the production of the dyon-antidyon pair via photon fusion for spins $0, 1/2$, and $1$, with magnetic charge $1g_D$ and electric charge varying from $1e$ to $200e$, in proton-proton collisions with a center-of-mass energy of 14~TeV.}
     \label{fig:combined-spins}
\end{figure}

We analyze the production of dyons by modifying certain parameters. In the Figures~(\ref{fig:spin-0}), (\ref{fig:spin-1/2}) and (\ref{fig:spin-1}), we present the behavior of the total cross section for the production of dyons with spin $0$, $1/2$ and $1$ as a function of the mass \(M_d\), fixing the magnetic charge number \(n_m=1\) and varying the electric charge number \(n_e=1\) to \(200\).

\begin{figure}[H]
    \centering
    \begin{subfigure}[b]{0.45\textwidth}
        \centering
        \includegraphics[width=\textwidth]{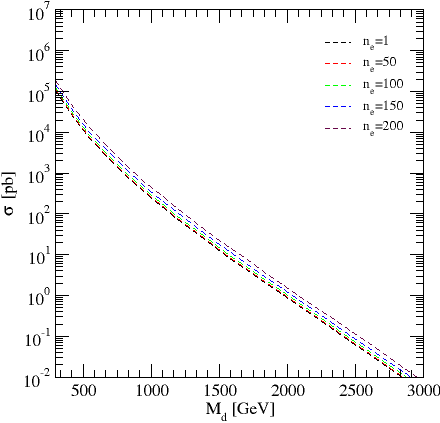}
        \caption{Spin 0}
        \label{fig:spins-0}
    \end{subfigure}
    \hfill
    \begin{subfigure}[b]{0.45\textwidth}
        \centering
        \includegraphics[width=\textwidth]{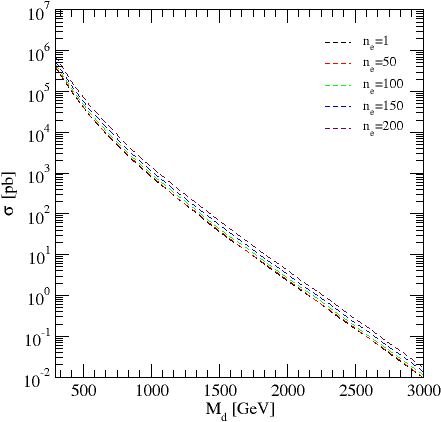}
        \caption{Spin $1/2$}
        \label{fig:spins-1/2}
    \end{subfigure}
    \hfill
    \begin{subfigure}[b]{0.45\textwidth}
        \centering
        \includegraphics[width=\textwidth]{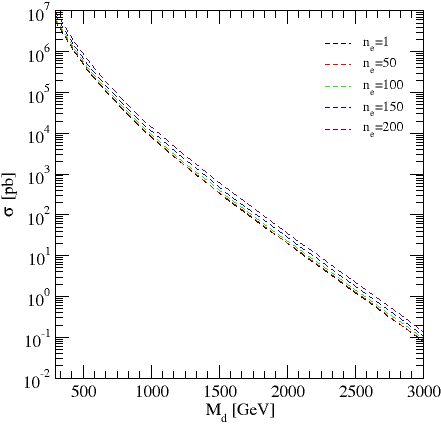}
        \caption{Spin 1}
        \label{fig:spins-1}
    \end{subfigure}

    \caption{Behavior of the cross section for the production of the dyon-antidyon pair via photon fusion for spins $0, 1/2$, and $1$, with magnetic charge $5g_D$ and electric charge varying from $1e$ to $200e$, in proton-proton collisions with a center-of-mass energy of 14~TeV.}
    \label{fig:combined-spins-ng5}
\end{figure}

In the Figure~(\ref{fig:spins-0}), (\ref{fig:spins-1/2}) and (\ref{fig:spins-1}), we present the behavior of the total cross section for the production of dyons with spin $0$, $1/2$, and $1$ as a function of the mass \(M_d\). In this case, the results are for a magnetic charge number \(n_m=5\), and similarly, we vary the electric charge number \(n_e=1\) to \(200\).

By comparing the results, we observe some subtle differences in the cross sections of dyons with different spins. For spin-$0$ dyons, the cross section is the smallest among the cases analyzed, which aligns with the fact that spin $0$ particles have fewer degrees of freedom to interact with other particles. For spin-$1/2$ dyons, the cross section is slightly larger than that of spin-$0$, suggesting a higher probability of production for dyons with a semi-integer spin. Finally, spin-$1$ dyons exhibit the largest cross section among the three cases, which can be explained by the greater number of interaction modes available for particles with higher spin, increasing the production in collisions.

In all the figures, the cross section decreases as the mass of the dyons increases, which is expected because heavier particles require more energy to be produced, resulting in a lower probability of production as the mass grows.

The lines in the figures (\ref{fig:combined-spins}), (\ref{fig:combined-spins-ng5})correspond to different values of the electric charge \(n_e\), ranging from 1 to 200. For \(n_m=1\), the cross section showed a significant increase as the value of \(n_e\) increased. For \(n_m=5\), the behavior of the cross section is quite similar, although the variations are less pronounced. In both cases, there is a clear dependence on the electric charge, with the cross section gradually increasing as \(n_e\) increased. This behavior can be explained by cross section equations~(\ref{7}), (\ref{4.11}), and (\ref{4.13}), which are directly proportional to the square of the fine structure constant (\(\sigma \propto \alpha_d^2\)), such that larger electric charges result in higher cross sections.

To connect the calculated cross sections with the conditions of the LHC during its third operational phase (Run-3), we provide a coarse estimate of the number of expected events ($N$), these bare estimates do not account for exclusivity survival factors ($S^2$), detector acceptances, or kinematic cuts (such as minimum transverse momentum or pseudorapidity constraints), which would be strictly required in realistic experimental setups like MoEDAL or ATLAS. This estimate is obtained using the relation $N = \sigma \cdot L_{int}$, where $L_{int}$ represents the integrated luminosity. In this analysis, we adopt a specific integrated luminosity of $\unit[160.4]{fb^{-1}}$ $(1.604 \times 10^5 \text{ pb}^{-1})$, which reflects the data accumulated in recent high-energy operational runs. For our estimations, we focused on dyons with spin-$1$, which exhibited the highest interaction probability among the cases analyzed. We assumed a magnetic charge number $n_m = 1$ and an electric charge $n_e = 1$. We provide a theoretical upper bound for the expected number of events.

Table (\ref{table1}) summarizes the expected yields for different mass scales at a center-of-mass energy of $\unit[14]{TeV}$.
\begin{table}[h]
	\centering
    \renewcommand{\arraystretch}{1.3}
	\caption{Expected number of events ($N$) for the production of spin-1 dyon-antidyon pairs through photon fusion, considering unit magnetic and electric charges ($n_m = 1, n_e = 1$) at $\sqrt{s} = 14$ TeV. The calculation assumes an integrated luminosity of $L_{int} = 160.4$ fb$^{-1}$.}
	\begin{tabular*}{\textwidth}{@{\extracolsep{\fill}}ccc@{}}
		\hline\hline
		Mass $(M_d)$ & Cross Section $(\sigma)$ & Expected Events ($N$) \\ \hline
		750 GeV & $\approx 8.30 \times 10^1$ pb & $\approx 1.33 \times 10^7$ \\
		1000 GeV & $\approx 1.31 \times 10^1$ pb & $\approx 2.10 \times 10^6$ \\
		1500 GeV & $\approx 5.53 \times 10^{-1}$ pb & $\approx 8.87 \times 10^4$ \\ 
        1910 GeV & $\approx 5.27 \times 10^{-2}$ pb & $\approx 8.46 \times 10^3$ \\ \hline\hline
	\end{tabular*}
    \label{table1}
\end{table}

Analysis of the data in Table (\ref{table1}) reveals the drastic impact of kinematic suppression. At the lower limit of the allowed window ($M_d = \unit[750]{GeV}$), the model predicts a colossal production on the order of $\sim 10^7$ events. Magnetically charged particles are characterized as highly ionizing (HIPs), leaving anomalous and intense traces in the detectors. The non-observation of a signal of such magnitude corroborates the robust exclusion for $M_d < \unit[750]{GeV}$.

As the mass increases to $\unit[1500]{GeV}$, the production rate suffers a reduction of more than two orders of magnitude ($N \approx 8.87 \times 10^4$), making the discrimination of the signal against the Standard Model background noise a substantial experimental challenge, dependent on the reconstruction efficiency and trigger criteria of the detectors.

Therefore, at the kinematic limit of $\unit[1910]{GeV}$, the number of raw events drops to $8.46 \times 10^3$. Considering that the acceptance efficiency of detectors for monopoles and dyons is typically restricted to small percentage fractions, this quantity of generated events translates into a marginal number of possible detections. This behavior justifies the $\unit[1910]{GeV}$ limit not as a theoretical prohibition, but as the statistical exhaustion point of the experimental sensitivity of Run-2, defining the primary target for future High-Luminosity LHC (HL-LHC) campaigns.

\section{Summary and Conclusions}

In this study, we investigated the production of dyons in proton-proton collisions via photon fusion at the LHC, considering dyons with spin $0$, $1/2$, and $1$, masses between $750$ and \unit[1910]{GeV}, magnetic charges ranging from \(1g_D\) to \(5g_D\), and electric charges varying from \(1e\) to \(200e\). In all the analyzed cases, we observed that the production of dyons was significant within the estimated energy range. We identified that the total cross section increases with both the spin and electric charge of the produced particle, indicating that dyons with a higher spin and greater electric charge have a higher probability of interaction in collisions.

Additionally, our results showed that the production of dyons was strongly suppressed for higher mass dyons, reflecting a significant decrease in the cross section. This behavior highlights the difficulties associated with the experimental observation of massive dyons, although lower mass particles may be more accessible in the current experiments. In comparison with the Drell-Yan mechanism used by MoEDAL, while DY involves quark-antiquark annihilation, photon fusion is an alternative channel where projectiles do not dissociate, providing a cleaner experimental signal in the central rapidity region. Possible sources of uncertainty include the modeling of the proton's internal structure and the form of the equivalent photon spectrum and the effective nature of the theory employed.

Throughout this study, we emphasize the relevance of dyons as a natural extension of magnetic monopoles and their potential contributions to the understanding of electric charge quantization and the unification of fundamental forces. The possible discovery of dyons would open a new window in particle physics, revealing crucial details regarding the nature of magnetic monopoles and electromagnetic interactions.

Although experiments conducted thus far, including those at the LHC, have not identified the presence of dyons, the continuation of this search is of utmost importance. The detection of these exotic particles represents a significant advancement beyond the Standard Model, providing insights into high-energy physics and exploring previously unknown phenomena. Thus, the study of dyon production in future experiments, especially under higher energy conditions, remains one of the most promising challenges and opportunities in the search for new physics.

\section{Acknowledgments}
The author thanks the Grupo de Altas e M\'{e}dias Energias (IFM-UFPel) for their support at all stages of this work. In\'{a}cio, L. F. C. thanks CAPES and Carrefour for financial support during the development of this work.

\section*{Funding}
In\'{a}cio, L. F. C. thanks Coordena\c{c}\~ao de Aperfei\c{c}oamento de Pessoal de N\'{\i}vel superior (CAPES), Brazil, Finance Code 001, and Carrefour for financial support (MSc Scholarship) during the development of this work.

\section*{Data Availability Statement}
The data produced for this article is available on reasonable request from the authors.



\bibliography{biblio}


\begin{thebibliography}{43}
\ifx \bisbn   \undefined \def \bisbn  #1{ISBN #1}\fi
\ifx \binits  \undefined \def \binits#1{#1}\fi
\ifx \bauthor  \undefined \def \bauthor#1{#1}\fi
\ifx \batitle  \undefined \def \batitle#1{#1}\fi
\ifx \bjtitle  \undefined \def \bjtitle#1{#1}\fi
\ifx \bvolume  \undefined \def \bvolume#1{\textbf{#1}}\fi
\ifx \byear  \undefined \def \byear#1{#1}\fi
\ifx \bissue  \undefined \def \bissue#1{#1}\fi
\ifx \bfpage  \undefined \def \bfpage#1{#1}\fi
\ifx \blpage  \undefined \def \blpage #1{#1}\fi
\ifx \burl  \undefined \def \burl#1{\textsf{#1}}\fi
\ifx \doiurl  \undefined \def \doiurl#1{\url{https://doi.org/#1}}\fi
\ifx \betal  \undefined \def \betal{\textit{et al.}}\fi
\ifx \binstitute  \undefined \def \binstitute#1{#1}\fi
\ifx \binstitutionaled  \undefined \def \binstitutionaled#1{#1}\fi
\ifx \bctitle  \undefined \def \bctitle#1{#1}\fi
\ifx \beditor  \undefined \def \beditor#1{#1}\fi
\ifx \bpublisher  \undefined \def \bpublisher#1{#1}\fi
\ifx \bbtitle  \undefined \def \bbtitle#1{#1}\fi
\ifx \bedition  \undefined \def \bedition#1{#1}\fi
\ifx \bseriesno  \undefined \def \bseriesno#1{#1}\fi
\ifx \blocation  \undefined \def \blocation#1{#1}\fi
\ifx \bsertitle  \undefined \def \bsertitle#1{#1}\fi
\ifx \bsnm \undefined \def \bsnm#1{#1}\fi
\ifx \bsuffix \undefined \def \bsuffix#1{#1}\fi
\ifx \bparticle \undefined \def \bparticle#1{#1}\fi
\ifx \barticle \undefined \def \barticle#1{#1}\fi
\bibcommenthead
\ifx \bconfdate \undefined \def \bconfdate #1{#1}\fi
\ifx \botherref \undefined \def \botherref #1{#1}\fi
\ifx \url \undefined \def \url#1{\textsf{#1}}\fi
\ifx \bchapter \undefined \def \bchapter#1{#1}\fi
\ifx \bbook \undefined \def \bbook#1{#1}\fi
\ifx \bcomment \undefined \def \bcomment#1{#1}\fi
\ifx \oauthor \undefined \def \oauthor#1{#1}\fi
\ifx \citeauthoryear \undefined \def \citeauthoryear#1{#1}\fi
\ifx \endbibitem  \undefined \def \endbibitem {}\fi
\ifx \bconflocation  \undefined \def \bconflocation#1{#1}\fi
\ifx \arxivurl  \undefined \def \arxivurl#1{\textsf{#1}}\fi
\csname PreBibitemsHook\endcsname

\bibitem[\protect\citeauthoryear{Schwinger}{1969}]{Schwinger:1969ib}
\begin{barticle}
\bauthor{\bsnm{Schwinger}, \binits{J.S.}}
\bjtitle{Science}
\bvolume{165},
\bfpage{757}--\blpage{761}
(\byear{1969})
\end{barticle}
\endbibitem

\bibitem[\protect\citeauthoryear{Mavromatos and
  Mitsou}{2020}]{Mavromatos:2020gwk}
\begin{barticle}
\bauthor{\bsnm{Mavromatos}, \binits{N.E.}},
\bauthor{\bsnm{Mitsou}, \binits{V.A.}}:
\batitle{{Magnetic monopoles revisited: Models and searches at colliders and in
  the Cosmos}}.
\bjtitle{Int. J. Mod. Phys. A}
\bvolume{35}(\bissue{23}),
\bfpage{2030012}
(\byear{2020})
\doiurl{10.1142/S0217751X20300124}
{\href{https://arxiv.org/abs/2005.05100}{{arXiv:2005.05100}}}
{[hep-ph]}
\end{barticle}
\endbibitem

\bibitem[\protect\citeauthoryear{Dirac}{1931}]{Dirac1931}
\begin{barticle}
\bauthor{\bsnm{Dirac}, \binits{P.A.M.}}
\bjtitle{Proc.Roy.Soc.Lond}
\bvolume{133},
\bfpage{60}--\blpage{72}
(\byear{1931})
\end{barticle}
\endbibitem

\bibitem[\protect\citeauthoryear{l'Yi et~al.}{1982}]{Dyon_GUT82}
\begin{barticle}
\bauthor{\bsnm{l'Yi}, \binits{W.S.}},
\bauthor{\bsnm{Park}, \binits{Y.J.}},
\bauthor{\bsnm{Koh}, \binits{I.G.}},
\bauthor{\bsnm{Kim}, \binits{Y.D.}}
\bjtitle{Physical Review Letters}
\bvolume{49},
\bfpage{1229}--\blpage{1231}
(\byear{1982})
\end{barticle}
\endbibitem

\bibitem[\protect\citeauthoryear{Zhong}{1984}]{Dyon_GUT84}
\begin{barticle}
\bauthor{\bsnm{Zhong}, \binits{Q.M.}}
\bjtitle{Nuclear Physics B}
\bvolume{231}(\bissue{1}),
\bfpage{172}--\blpage{188}
(\byear{1984})
\end{barticle}
\endbibitem

\bibitem[\protect\citeauthoryear{Bjoraker and
  Hosotani}{2000}]{Dyon_Einstein-Yang-Mills2000}
\begin{barticle}
\bauthor{\bsnm{Bjoraker}, \binits{J.}},
\bauthor{\bsnm{Hosotani}, \binits{Y.}}
\bjtitle{Physical Review Letters}
\bvolume{84}(\bissue{9}),
\bfpage{1853}
(\byear{2000})
\end{barticle}
\endbibitem

\bibitem[\protect\citeauthoryear{Nolan and
  Winstanley}{2012}]{Dyon_Einstein-Yang-Mills2012}
\begin{barticle}
\bauthor{\bsnm{Nolan}, \binits{B.C.}},
\bauthor{\bsnm{Winstanley}, \binits{E.}}
\bjtitle{Classical and Quantum Gravity}
\bvolume{29}(\bissue{23}),
\bfpage{235024}
(\byear{2012})
\end{barticle}
\endbibitem

\bibitem[\protect\citeauthoryear{Sen}{1997}]{Dyons_Kaluza-Klein1997}
\begin{barticle}
\bauthor{\bsnm{Sen}, \binits{A.}}
\bjtitle{Physical Review Letters}
\bvolume{79}(\bissue{9}),
\bfpage{1619}
(\byear{1997})
\end{barticle}
\endbibitem

\bibitem[\protect\citeauthoryear{Dabholkar
  et~al.}{2011}]{Dyons_Kaluza-Klein2011}
\begin{barticle}
\bauthor{\bsnm{Dabholkar}, \binits{A.}},
\bauthor{\bsnm{Gomes}, \binits{J.}},
\bauthor{\bsnm{Murthy}, \binits{S.}}
\bjtitle{Journal of High Energy Physics}
\bvolume{2011}(\bissue{5}),
\bfpage{1}--\blpage{28}
(\byear{2011})
\end{barticle}
\endbibitem

\bibitem[\protect\citeauthoryear{Gomez and Manjarin}{2001}]{Dyons_teoriaM2001}
\begin{botherref}
\oauthor{\bsnm{Gomez}, \binits{C.}},
\oauthor{\bsnm{Manjarin}, \binits{J.J.}}
ArXiv Preprint hep-th/0111169
(2001)
\end{botherref}
\endbibitem

\bibitem[\protect\citeauthoryear{Acharya et~al.}{2014}]{MoEDAL:2014ttp}
\begin{barticle}
\bauthor{\bsnm{Acharya}, \binits{B.}}, \betal
\bjtitle{International Journal of Modern Physics A}
\bvolume{29},
\bfpage{1430050}
(\byear{2014})
{[hep-ph]}
\end{barticle}
\endbibitem

\bibitem[\protect\citeauthoryear{Pinfold et~al.}{2009}]{MoEDAL1}
\begin{botherref}
\oauthor{\bsnm{Pinfold}, \binits{J.}}, et al.
MoEDAL Collaboration
(2009)
\end{botherref}
\endbibitem

\bibitem[\protect\citeauthoryear{Acharya et~al.}{2021}]{primeirabusca}
\begin{barticle}
\bauthor{\bsnm{Acharya}, \binits{B.}}, \betal
\bjtitle{Physical Review Letters}
\bvolume{126},
\bfpage{071801}
(\byear{2021})
\end{barticle}
\endbibitem

\bibitem[\protect\citeauthoryear{Bertulani et~al.}{2005}]{bertulani2005physics}
\begin{barticle}
\bauthor{\bsnm{Bertulani}, \binits{C.A.}},
\bauthor{\bsnm{Klein}, \binits{S.R.}},
\bauthor{\bsnm{Nystrand}, \binits{J.}}, \betal:
\batitle{Physics of ultra-peripheral nuclear collisions}.
\bjtitle{Annu. Rev. Nucl. Part. Sci.}
\bvolume{55}(\bissue{1}),
\bfpage{271}--\blpage{310}
(\byear{2005})
\end{barticle}
\endbibitem

\bibitem[\protect\citeauthoryear{Baltz et~al.}{2008}]{baltz2008physics}
\begin{barticle}
\bauthor{\bsnm{Baltz}, \binits{A.J.}},
\bauthor{\bsnm{Baur}, \binits{G.}},
\bauthor{\bsnm{d{\rq}Enterria}, \binits{D.}},
\bauthor{\bsnm{Frankfurt}, \binits{L.}},
\bauthor{\bsnm{Gelis}, \binits{F.}},
\bauthor{\bsnm{Guzey}, \binits{V.}},
\bauthor{\bsnm{Hencken}, \binits{K.}},
\bauthor{\bsnm{Kharlov}, \binits{Y.}},
\bauthor{\bsnm{Klasen}, \binits{M.}},
\bauthor{\bsnm{Klein}, \binits{S.R.}}:
\batitle{The physics of ultraperipheral collisions at the lhc}.
\bjtitle{Physics Reports}
\bvolume{458}(\bissue{1-3}),
\bfpage{1}--\blpage{171}
(\byear{2008})
\end{barticle}
\endbibitem

\bibitem[\protect\citeauthoryear{Drees and
  Zeppenfeld}{1989}]{Drees_Zeppenfeld1989}
\begin{barticle}
\bauthor{\bsnm{Drees}, \binits{M.}},
\bauthor{\bsnm{Zeppenfeld}, \binits{D.}}:
\batitle{Production of supersymmetric particles in elastic ep collisions}.
\bjtitle{Physical Review D}
\bvolume{39},
\bfpage{2536}--\blpage{2546}
(\byear{1989})
\end{barticle}
\endbibitem

\bibitem[\protect\citeauthoryear{Acharya et~al.}{2019}]{mono2}
\begin{barticle}
\bauthor{\bsnm{Acharya}, \binits{B.}},
\bauthor{\bsnm{Alexandre}, \binits{J.}},
\bauthor{\bsnm{Baines}}, \betal
\bjtitle{Physical Review Letters}
\bvolume{123},
\bfpage{021802}
(\byear{2019})
\end{barticle}
\endbibitem

\bibitem[\protect\citeauthoryear{Acharya et~al.}{2018}]{mono1}
\begin{barticle}
\bauthor{\bsnm{Acharya}, \binits{B.}},
\bauthor{\bsnm{Alexandre}, \binits{J.}},
\bauthor{\bsnm{Baines}, \binits{S.}}, \betal
\bjtitle{Physics Letters B}
\bvolume{782},
\bfpage{510}--\blpage{516}
(\byear{2018})
\end{barticle}
\endbibitem

\bibitem[\protect\citeauthoryear{Alwall et~al.}{2014}]{Alwall:2014hca}
\begin{barticle}
\bauthor{\bsnm{Alwall}, \binits{J.}},
\bauthor{\bsnm{Frederix}, \binits{R.}},
\bauthor{\bsnm{Frixione}, \binits{S.}},
\bauthor{\bsnm{others.}}
\bjtitle{Journal of High Energy Physics}
\bvolume{07},
\bfpage{079}
(\byear{2014})
\end{barticle}
\endbibitem

\bibitem[\protect\citeauthoryear{Ball et~al.}{2013}]{Ball2012}
\begin{barticle}
\bauthor{\bsnm{Ball}, \binits{R.D.}}, \betal
\bjtitle{Nuclear Physics B}
\bvolume{867},
\bfpage{244}--\blpage{289}
(\byear{2013})
\end{barticle}
\endbibitem

\bibitem[\protect\citeauthoryear{Ennadifi}{2024}]{ennadifi2024light}
\begin{barticle}
\bauthor{\bsnm{Ennadifi}, \binits{S.E.}}:
\batitle{On light-by-light interaction in qed}.
\bjtitle{International Journal of Theoretical Physics}
\bvolume{63}(\bissue{9}),
\bfpage{1}--\blpage{9}
(\byear{2024})
\end{barticle}
\endbibitem

\bibitem[\protect\citeauthoryear{d'Enterria and
  da~Silveira}{2013}]{dEnterria2013}
\begin{barticle}
\bauthor{\bsnm{d'Enterria}, \binits{D.}},
\bauthor{\bsnm{Silveira}, \binits{G.G.}}
\bjtitle{Phys. Rev. Lett.}
\bvolume{111},
\bfpage{080405}
(\byear{2013})
\end{barticle}
\endbibitem

\bibitem[\protect\citeauthoryear{Klasen}{2019}]{klasen2019perspectives}
\begin{barticle}
\bauthor{\bsnm{Klasen}, \binits{M.}}:
\batitle{{Perspectives of photon physics at future colliders}}.
\bjtitle{Frascati Phys. Ser.}
\bvolume{69},
\bfpage{7}--\blpage{17}
(\byear{2019})
\end{barticle}
\endbibitem

\bibitem[\protect\citeauthoryear{Belforte et~al.}{2019}]{belforte2019evidence}
\begin{barticle}
\bauthor{\bsnm{Belforte}, \binits{S.}},
\bauthor{\bsnm{Candelise}, \binits{V.}},
\bauthor{\bsnm{Casarsa}, \binits{M.}},
\bauthor{\bsnm{{DA ROLD}}, \binits{A.}},
\bauthor{\bsnm{Vazzoler}, \binits{F.}},
\bauthor{\bsnm{Zanetti}, \binits{A.}}:
\batitle{Evidence for light-by-light scattering and searches for axion-like
  particles in ultraperipheral pbpb collisions at sqrt (s\_nn)= 5.02 tev}.
\bjtitle{Physics Letters B}
\bvolume{797},
\bfpage{1}--\blpage{27}
(\byear{2019})
\end{barticle}
\endbibitem

\bibitem[\protect\citeauthoryear{Harland et~al.}{2016}]{harland2016production}
\begin{barticle}
\bauthor{\bsnm{Harland}, \binits{L.A.}},
\bauthor{\bsnm{Khoze}, \binits{V.A.}},
\bauthor{\bsnm{Ryskin}, \binits{M.G.}}:
\batitle{The production of a diphoton resonance via photon-photon fusion}.
\bjtitle{Journal of High Energy Physics}
\bvolume{2016}(\bissue{3}),
\bfpage{1}--\blpage{27}
(\byear{2016})
\end{barticle}
\endbibitem

\bibitem[\protect\citeauthoryear{Silveira et~al.}{2015}]{da2015central}
\begin{barticle}
\bauthor{\bsnm{Silveira}, \binits{G.G.}},
\bauthor{\bsnm{Forthomme}, \binits{L.}},
\bauthor{\bsnm{Piotrzkowski}, \binits{K.}},
\bauthor{\bsnm{Sch{\"a}fer}, \binits{W.}},
\bauthor{\bsnm{Szczurek}, \binits{A.}}:
\batitle{Central $\mu$+ $\mu$- production via photon-photon fusion in
  proton-proton collisions with proton dissociation}.
\bjtitle{Journal of High Energy Physics}
\bvolume{2015}(\bissue{2}),
\bfpage{1}--\blpage{24}
(\byear{2015})
\end{barticle}
\endbibitem

\bibitem[\protect\citeauthoryear{d{\rq}Enterria and
  Shao}{2022}]{d2022observing}
\begin{barticle}
\bauthor{\bsnm{d{\rq}Enterria}, \binits{D.}},
\bauthor{\bsnm{Shao}, \binits{H.S.}}:
\batitle{Observing true tauonium via two-photon fusion at e+ e-and hadron
  colliders}.
\bjtitle{Physical Review D}
\bvolume{105}(\bissue{9}),
\bfpage{093008}
(\byear{2022})
\end{barticle}
\endbibitem

\bibitem[\protect\citeauthoryear{Chiesa et~al.}{2024}]{chiesa2024multi}
\begin{barticle}
\bauthor{\bsnm{Chiesa}, \binits{M.}},
\bauthor{\bsnm{Mele}, \binits{B.}},
\bauthor{\bsnm{Piccinini}, \binits{F.}}:
\batitle{Multi higgs production via photon fusion at future multi-tev muon
  colliders}.
\bjtitle{The European Physical Journal C}
\bvolume{84}(\bissue{5}),
\bfpage{1}--\blpage{17}
(\byear{2024})
\end{barticle}
\endbibitem

\bibitem[\protect\citeauthoryear{Vidovi{\'c} et~al.}{1993}]{vidovic1993impact}
\begin{barticle}
\bauthor{\bsnm{Vidovi{\'c}}, \binits{M.}},
\bauthor{\bsnm{Greiner}, \binits{M.}},
\bauthor{\bsnm{Best}, \binits{C.}},
\bauthor{\bsnm{Soff}, \binits{G.}}:
\batitle{Impact-parameter dependence of the electromagnetic particle production
  in ultrarelativistic heavy-ion collisions}.
\bjtitle{Physical Review C}
\bvolume{47}(\bissue{5}),
\bfpage{2308}
(\byear{1993})
\end{barticle}
\endbibitem

\bibitem[\protect\citeauthoryear{{\c S}ahin and Inan}{2009}]{csahin2009probe}
\begin{barticle}
\bauthor{\bsnm{{\c S}ahin}, \binits{{\. I}.}},
\bauthor{\bsnm{Inan}, \binits{S.C.}}:
\batitle{Probe of unparticles at the lhc in exclusive two lepton and two photon
  production via photon-photon fusion}.
\bjtitle{Journal of High Energy Physics}
\bvolume{2009}(\bissue{09}),
\bfpage{069}
(\byear{2009})
\end{barticle}
\endbibitem

\bibitem[\protect\citeauthoryear{Dougall and Wick}{2009}]{dougall2009dirac}
\begin{barticle}
\bauthor{\bsnm{Dougall}, \binits{T.}},
\bauthor{\bsnm{Wick}, \binits{S.D.}}:
\batitle{Dirac magnetic monopole production from photon fusion in proton
  collisions}.
\bjtitle{The European Physical Journal A}
\bvolume{39}(\bissue{2}),
\bfpage{213}--\blpage{217}
(\byear{2009})
\end{barticle}
\endbibitem

\bibitem[\protect\citeauthoryear{Szczurek}{2002}]{szczurek2002heavy}
\begin{barticle}
\bauthor{\bsnm{Szczurek}, \binits{A.}}:
\batitle{Heavy quark production in photon-nucleon and photon-photon
  collisions}.
\bjtitle{The European Physical Journal C-Particles and Fields}
\bvolume{26}(\bissue{2}),
\bfpage{183}--\blpage{194}
(\byear{2002})
\end{barticle}
\endbibitem

\bibitem[\protect\citeauthoryear{Bertulani et~al.}{2025}]{Bertulani:2024vpt}
\begin{barticle}
\bauthor{\bsnm{Bertulani}, \binits{C.A.}},
\bauthor{\bsnm{Francener}, \binits{R.}},
\bauthor{\bsnm{Gon{\c c}alves}, \binits{V.P.}},
\bauthor{\bsnm{Souza}, \binits{J.T.}}:
\batitle{Particle production by $\gamma$-$\gamma$ interactions in future
  electron-ion colliders}.
\bjtitle{Physical Review C}
\bvolume{111}(\bissue{2}),
\bfpage{025201}
(\byear{2025})
\end{barticle}
\endbibitem

\bibitem[\protect\citeauthoryear{Mikaelian}{1982}]{Mikaelian1981}
\begin{barticle}
\bauthor{\bsnm{Mikaelian}, \binits{K.O.}}
\bjtitle{Phys. Lett. B}
\bvolume{115}(\bissue{3}),
\bfpage{267}--\blpage{269}
(\byear{1982})
\end{barticle}
\endbibitem

\bibitem[\protect\citeauthoryear{Moulin et~al.}{1996}]{Moulin1996}
\begin{barticle}
\bauthor{\bsnm{Moulin}, \binits{F.}},
\bauthor{\bsnm{Bernard}, \binits{D.}},
\bauthor{\bsnm{Amiranoff}, \binits{F.}}
\bjtitle{Z. Phys. C}
\bvolume{72},
\bfpage{607}--\blpage{611}
(\byear{1996})
\end{barticle}
\endbibitem

\bibitem[\protect\citeauthoryear{Jikia and Tkabladze}{1994}]{Jikia1993}
\begin{barticle}
\bauthor{\bsnm{Jikia}, \binits{G.}},
\bauthor{\bsnm{Tkabladze}, \binits{A.}}
\bjtitle{Phys. Lett. B}
\bvolume{323},
\bfpage{453}--\blpage{458}
(\byear{1994})
\end{barticle}
\endbibitem

\bibitem[\protect\citeauthoryear{d'Enterria et~al.}{2008}]{dEnterria2008p}
\begin{barticle}
\bauthor{\bsnm{d'Enterria}, \binits{D.}},
\bauthor{\bsnm{Klasen}, \binits{M.}},
\bauthor{\bsnm{Piotrzkowski}, \binits{K.}}
\bjtitle{Nucl. Phys. B Proc. Suppl.}
\bvolume{179},
\bfpage{1}
(\byear{2008})
\end{barticle}
\endbibitem

\bibitem[\protect\citeauthoryear{Gamberg and Milton}{2000a}]{Gamberg2000}
\begin{barticle}
\bauthor{\bsnm{Gamberg}, \binits{L.}},
\bauthor{\bsnm{Milton}, \binits{K.A.}}
\bjtitle{Physical Review D}
\bvolume{61},
\bfpage{075013}
(\byear{2000})
\end{barticle}
\endbibitem

\bibitem[\protect\citeauthoryear{Gamberg and Milton}{2000b}]{Gamberg:2000}
\begin{botherref}
\oauthor{\bsnm{Gamberg}, \binits{L.P.}},
\oauthor{\bsnm{Milton}, \binits{K.A.}}:
In: 5th Workshop on QCD (QCD 2000),
pp. 176--185
(2000)
\end{botherref}
\endbibitem

\bibitem[\protect\citeauthoryear{Baines et~al.}{2018}]{Baines:2018ltl}
\begin{barticle}
\bauthor{\bsnm{Baines}, \binits{S.}},
\bauthor{\bsnm{Mavromatos}, \binits{N.E.}},
\bauthor{\bsnm{Mitsou}, \binits{V.A.}},
\bauthor{\bsnm{Pinfold}, \binits{J.L.}},
\bauthor{\bsnm{Santra}, \binits{A.}}
\bjtitle{Eur. Phys. J. C}
\bvolume{78}(\bissue{11}),
\bfpage{966}
(\byear{2018})
{[hep-ph]}
\end{barticle}
\endbibitem

\bibitem[\protect\citeauthoryear{Kurochkin et~al.}{2006}]{kurochkin2006}
\begin{barticle}
\bauthor{\bsnm{Kurochkin}, \binits{Y.}},
\bauthor{\bsnm{Satsunkevich}, \binits{I.}},
\bauthor{\bsnm{Shoukavy}, \binits{D.}},
\bauthor{\bsnm{Rusakovich}, \binits{N.}},
\bauthor{\bsnm{Kulchitsky}, \binits{Y.}}
\bjtitle{Modern Physics Letters A}
\bvolume{21}(\bissue{38}),
\bfpage{2873}--\blpage{2880}
(\byear{2006})
\end{barticle}
\endbibitem

\bibitem[\protect\citeauthoryear{Reis and Sauter}{2017}]{Reis:2017rvb}
\begin{barticle}
\bauthor{\bsnm{Reis}, \binits{J.T.}},
\bauthor{\bsnm{Sauter}, \binits{W.K.}}:
\batitle{{Production of magnetic monopoles and monopolium in peripheral
  collisions}}.
\bjtitle{Phys. Rev. D}
\bvolume{96}(\bissue{7}),
\bfpage{075031}
(\byear{2017})
\doiurl{10.1103/PhysRevD.96.075031}
{\href{https://arxiv.org/abs/1707.04170}{{arXiv:1707.04170}}}
{[hep-ph]}
\end{barticle}
\endbibitem

\bibitem[\protect\citeauthoryear{da~Silva and Sauter}{2024}]{daSilva:2023jxd}
\begin{barticle}
\bauthor{\bsnm{Silva}, \binits{J.V.B.}},
\bauthor{\bsnm{Sauter}, \binits{W.K.}}:
\batitle{{Production of bound states of magnetic monopoles in high-energy
  collisions at LHC}}.
\bjtitle{Int. J. Mod. Phys. A}
\bvolume{39}(\bissue{11n12}),
\bfpage{2450048}
(\byear{2024})
\doiurl{10.1142/S0217751X24500489}
{\href{https://arxiv.org/abs/2308.15587}{{arXiv:2308.15587}}}
{[hep-ph]}
\end{barticle}
\endbibitem

\end{thebibliography}

\end{document}